\documentclass[aps,prl,twocolumn,showpacs]{revtex4-1}
\usepackage{graphicx}
\usepackage{subfigure}
\usepackage{amsmath,amssymb}
\usepackage{color}
\usepackage[usenames,dvipsnames,svgnames,table]{xcolor}
\usepackage{url}

\newcommand{\fref}[1]{Fig.~\ref{#1}} %

\begin{document}
\title{Novel modulated phase of liquid crystals: Covariant elasticity in the context of soft, achiral smectic-$C$ materials}
\author{Venkatramanan P.R.}
\author{Yashodhan Hatwalne}
\author{N.V. Madhusudana}
\affiliation{Raman Research Institute, C.V. Raman Avenue, Bangalore 560 080, India}

\date{\today}

\begin{abstract}
Ginzburg-Landau-de Gennes -type covariant theories are extensively used in connection with twist grain boundary (TGB) phases of \textit{chiral} smectogens. We analyze the stability conditions for the linear, covariant elasticity theory of smectic-$C$ liquid crystals in the context of \textit{achiral} materials, and predict an equilibrium modulated structure with an oblique wavevector. We suggest that a previous experimental observation of stripes in smectic-$C$ is consistent with the predicted structure.
\end{abstract}

\maketitle

Smectic liquid crystals (smectics) are one-dimensional  ``solids" composed of fluid layers exhibiting quasi-long-range periodic order along the layer normal \cite{deGP,CL}. de Gennes \cite{deG} recognized the close analogy between normal metal-superconductor and nematic-smectic-$A$ transitions. Superconductors as well as smectics are characterized by complex order parameters, and rotational invariance in smectics is the analog of gauge invariance in superconductors. Subsequent prediction of the detailed structure of the Abrikosov phase in type-II, \textit{chiral}  smectogenic materials \cite{RL}, and almost concurrent discovery \cite{Goodby} of this twist grain boundary (TGB) phase put the analogy on a firm footing. In accord with superconductors, smectics are classified as type-II (or type-I) depending on whether the Ginzburg parameter $ \kappa = \lambda/\xi > 1/\sqrt{2} $ (or $ < 1/\sqrt{2} $), where $ \lambda $ is the twist penetration depth (London penetration depth in superconductors) and $ \xi $ is the appropriate coherence length \cite{RL}. In TGB phases, intrinsic molecular chirality is the analog of external magnetic field in superconductors. 

\begin{figure}[htbp]
\includegraphics[scale=0.65]{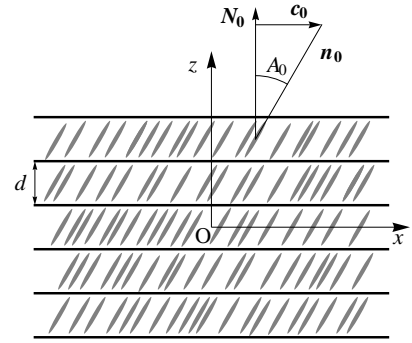}
\caption{ Schematic of Sm$C$. The $ xz $- plane is a mirror plane and $ O $ is a point of inversion. $ \mathbf{n}_{0} \equiv - \mathbf{n}_{0} $ is the unit Frank director, and the unit layer normal $ \mathbf{N}_{0} $ is along the $z$-axis. The polar vector $ \mathbf{c}_{0} = (c_{0}, 0, 0) $ is the projection of $ \mathbf{n}_{0} $ onto the plane of the layers. The equilibrium layer spacing is $d$. In Sm$A$, $\angle A_0 = 0$ and the layered structure is uniaxial.}
\label{FIG1}
\end{figure}

The class of smectic liquid crystals incorporates structures with diverse symmetry groups \cite{deGP}, such as that of smectic-$C$ (Sm$C$), which has a tilted director (\fref{FIG1}). This allows for a few TGB$_{C}$ phases which have been discovered experimentally and studied theoretically \cite{Barois}. All TGB phases have been found in chiral, type-II smectics, and are riddled with topological defects. For example, screw dislocations in the TGB$_{A}$ phase are analogs of flux tubes in the Abrikosov lattice. Theoretical investigations of TGB phases rely on covariant formulations of free energy  (based upon the superconductor-smectic analogy). Molecular chirality is modeled by including a term corresponding to the twist deformation of the director in the free energy.  \textit{TGB phases are not feasible in achiral materials, even if they are type-II in character}. 

\begin{figure}[htbp]
\includegraphics[scale=0.5]{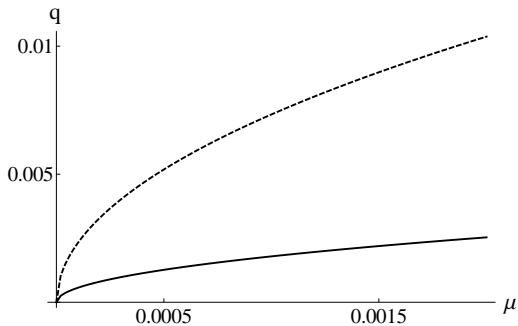}
\caption{The components of modulation wave-vector (measured in units of $ \xi^{-1} $), $ q_{x} $ (continuous) and $ q_{z} $ (dashed), as functions of $ \mu \equiv (l_{B}^{2}/d_{B}^{2} -1) $, for $ \kappa_{c} = 3 $, $ \kappa_{u} =1$ and $  d_{B}=1/3 $. For these parameter values, the stability range of the modulated phase  is $ 0.1 \lesssim l_{B}^{2} \lesssim 0.25 $.}
\label{FIG2}
\end{figure}

\begin{figure}[t]
  \subfigure[$\,$Layer structure.]{\label{FIG3a}\includegraphics[scale=0.45]{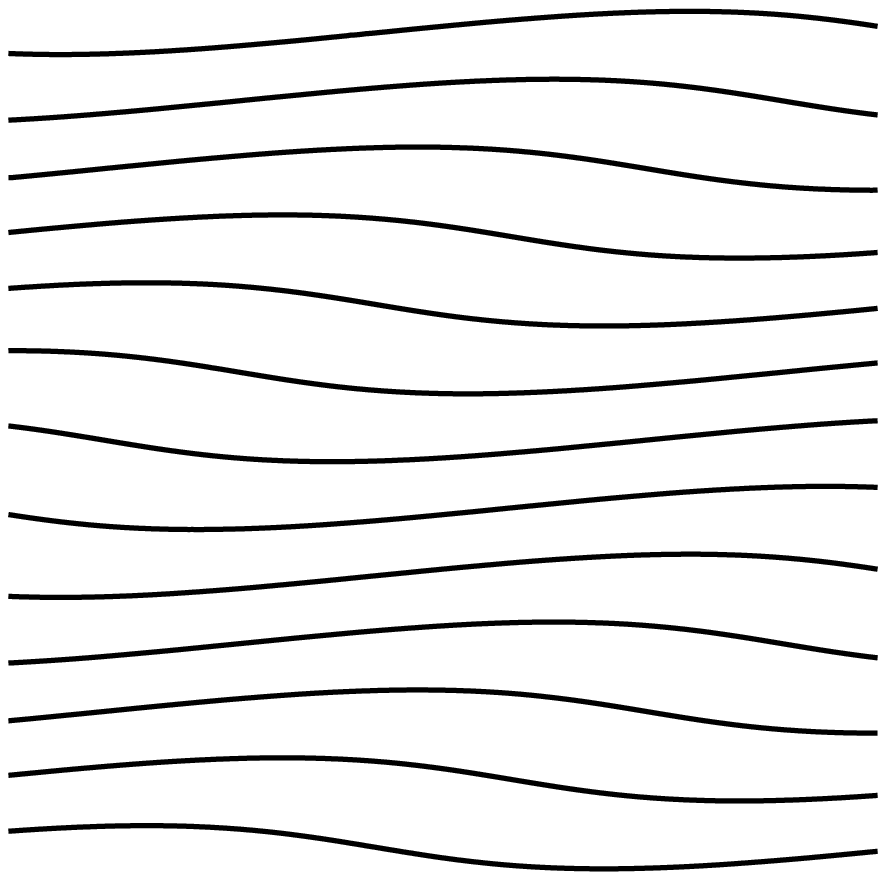}}
  \subfigure[$\,$Director field $\mathbf{n}$.]{\label{FIG3b}\includegraphics[scale=0.45]{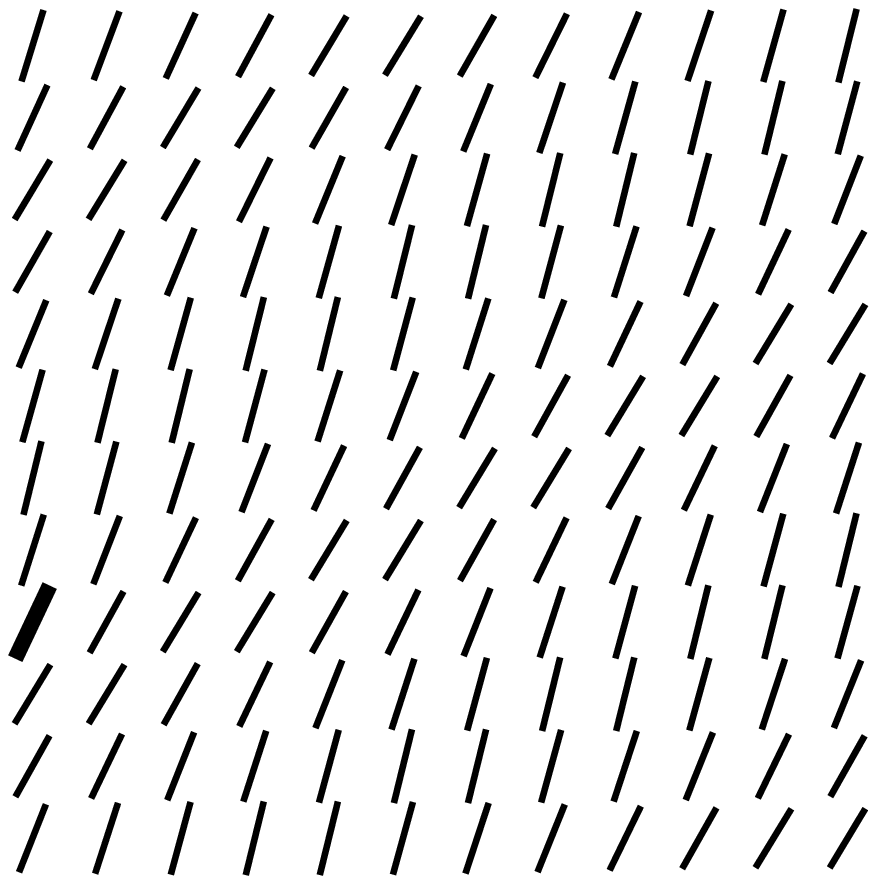}}
\caption{Schematic (exaggerated) of the modulated phase. The bold segment in (b) shows director orientation in undistorted Sm$C$. In the experiment discussed (see the text), the polarizer was placed along this direction, and the analyzer orthogonal to it. \textit{Although there is no translational order within the layers, line segments are placed periodically to emphasize the stripe structure}.}
\label{FIG3}
\end{figure}

Covariant \textit{elasticity} theories of smectics are obtained as ``low-temperature" limits (in which the modulus of the complex order parameter is fixed) of the Ginzburg-Landau-deGennes theory \cite{deG,RL,CL} and generalizations thereof \cite{KL}. These have almost exclusively been used in the context of type-II chiral smectogens. The modulated instability proposed in this letter is a manifestation of covariance in the unusual setting of \textit{achiral} Sm$C$ materials. We show that the linear, covariant elasticity theory of Sm$C$ \cite{HL} admits a transition to a modulated structure with an oblique wavevector (in the $ xz $- plane) as the ground state of the smectic medium (see \fref{FIG3}). Tilt order, the distinctive feature of Sm$C$, introduces new elastic couplings (that are absent in Sm$A$) in the covariant elasticity theory. Modulated instability sets in if the elastic constants discussed below satisfy the inequality $ L^{2} > B \, D $. Here $ B $ is the layer compression modulus, $ D $ is the coefficient of the covariant term which ensures that deviations from simultaneous, global rigid rotations of the layer normal with the Frank director cost energy, and $ L $ is the coefficient of the term that couples these two distortions in the elastic free energy (see the discussion following (\ref{f})). Tilt order is essential for the instability; symmetry of Sm$A$ prohibits a term analogous to the $ L $- term in the elastic free energy. Previous formulations of Sm$C$ elasticity, coupling molecular tilt to layer displacement (see, \textit{e.g.}, \cite{deGP}), do not lead to the instability discussed in this letter. Our results are summarized in Figs. \ref{FIG2} and \ref{FIG3}. We point out that an earlier observation of a periodic pattern in a Sm$C$ material \cite{PHM} is consistent with the predicted structure (see the discussion of results given below). 

Sm$C$ is a biaxial phase (\fref{FIG1}) in which the molecular director $ \mathbf{n}_{0} $ is tilted with respect to the layer normal $ \mathbf{N}_{0} $ so that $ \mathbf{n}_{0} \cdot \mathbf{N}_{0} = \cos A_{0} \equiv \alpha $. The projection of the Frank  director onto the layers is denoted by $ \mathbf{c}_{0} $. The plane spanned by $ \mathbf{n}_{0} $ and $ \mathbf{N}_{0} $ is a mirror plane with a center of inversion, and the structure is invariant under the simultaneous transformation $ \mathbf{N}_{0} \rightarrow -  \mathbf{N}_{0} $, $ \mathbf{c}_{0} \rightarrow -  \mathbf{c}_{0} $. In the distorted Sm$C$ the director $ \mathbf{n} = \mathbf{n}_{0} + \delta \mathbf{n} = \mathbf{c} + \sqrt{1 - c^{2}} \,\, \mathbf{N} $, 
where $ {\mathbf c}  = (c_{0} + \delta c) (\cos \phi, \sin \phi, 0) $, $ \delta c $ is the change in the magnitude of $ \mathbf{c} $, and $ \phi $ is the azimuthal angle. To the lowest order, the distortion in the director field is
\begin{equation}
\label{deltan}
\delta \mathbf{n} \simeq ( \delta c, \, c_{0} \, \delta \phi, \, - c_{0} \, \delta c/\sqrt{1 - c_{0}^{2} } ),
\end{equation}
and the distorted layer normal is given by
\begin{equation}
\label{N}
\mathbf{N} \simeq (- \partial_{x} u, -\partial_{y} u, 1),
\end{equation}
where the field $ u(\mathbf{x}) = u(x, y, z) $ measures the displacement of layers along the $ z $- direction. 
The broken symmetry elastic variables are $ u(\mathbf{x}) $, $ \delta c(\mathbf{x}) $ and $ \delta \phi(\mathbf{x}) $.  
The covariant, harmonic elastic free energy density is 
\cite{HL}
\begin{align}
\label{f}
f  = (1/2) \,  &[ B \, (\partial_{z} u)^{2} +  D \, (\delta c + \alpha \, \partial_{x} u)^{2} \nonumber \\
                  &- 2 L \, (\delta c + \alpha \, \partial_{x} u)(\partial_{z} u) + K_{u} \, (\nabla^{2} u)^{2} \nonumber \\
                  &+ K_{c} \, (\mathbf{\nabla} \delta c)^{2} + K_{\phi} \, (\mathbf{\nabla} \delta \phi)^{2} ],
\end{align}
with the elastic free energy functional given by $ F[u(\mathbf{x}), \delta c(\mathbf{x}), \delta \phi(\mathbf{x})] = \int f \, \mathrm{d}^{3} x $. The terms with coefficients $ B $ and $ K_{u} $ account for layer- compression and layer- bend energies respectively. In principle, the symmetry of Sm$C$ allows for anisotropy in the bend modulus. In (\ref{f}) above, the anisotropy in the effective bend modulus is taken into account indirectly via the coupling to the $ \delta c $- field (see \cite{HL} for a detailed discussion). The Frank free energy for distortions in the director field is represented via terms with coefficients $ K_{c} $ and $ K_{\phi} $ in the one-constant approximation \cite{deGP}. The $ D $- and $ L $- terms involve the form $ (\delta c + \alpha \, \partial_{x} u) $ which  measures the deviation  $ \delta (\mathbf{n} \cdot \mathbf{N}) = \mathbf{n}_{0} \cdot \delta \mathbf{N} +  \delta\mathbf{n} \cdot  \mathbf{N}_{0} $ from its equilibrium value $ \alpha $.  For simultaneous, rigid rotations of the Frank director and the layer normal, $ \delta (\mathbf{n} \cdot \mathbf{N}) = 0 $. The term with the coefficient $ L $ is allowed by the symmetry of Sm$C$, and couples $  \delta (\mathbf{n} \cdot \mathbf{N}) $ to changes in the equilibrium layer spacing.  This term is crucial for the proposed instability, and has no counterpart in the covariant elasticity theory of Sm$A$ \cite{CL}. The elastic constants $ B, D, K_{u}, K_{c} $ and $ K_{\phi} $ have to be positive for stability. Stability conditions do not restrict the sign of $ L $. Notice that the $ B $-, $ D$-, and $ L $ terms involve only $ x $- and $ z $- gradients of the broken symmetry variables. Furthermore, the $ \delta \phi $-  field is not coupled to the $ u $- and $ \delta c $- fields. The term with coefficient $ K_{\phi} $ plays no role in the modulated instability presented in this letter (see the discussion following (\ref{Fav})), and will be ignored in the following analysis. 

We now recast the elastic free energy in a form which is suited for the analysis of the proposed instability. In Fourier space the elastic free energy can be expressed as 
\begin{equation}
\label{FFourier}
F=\frac{1}{2}\int  \frac{\mathrm{d}^{3} q}{(2\pi)^{3}} \, \Phi_{a}^{\ast}(\mathbf{q}) \, G^{-1}_{ab} (\mathbf{q}) \, \Phi_b(\mathbf{q})  ,
\end{equation}
where repeated indices are summed over,  
$ \Phi_{1}(\mathbf{q}) = u(\mathbf{q}) $, $ \Phi_{2}(\mathbf{q}) = \delta c(\mathbf{q}) $, and
\begin{align}
\label{Gcomp}
G^{-1}_{11}(\mathbf{q}) &= B \, q_{z}^{2} + \alpha^{2} D \, q_{x}^{2} - 2 \alpha L \, q_{x} q_{z} + K_{u} \, q^{4},  \nonumber \\
G^{-1}_{12}(\mathbf{q}) &= - \, G^{-1}_{21}(\mathbf{q}) = i\, (L \, q_{z} -\alpha D \, q_{x}), \nonumber \\
G^{-1}_{22}(\mathbf{q}) &= D + K_{c} \, q^{2}.
\end{align} 
The Euler-Lagrange equations are
\begin{equation}
\label{EL}
\frac{\delta F}{\delta \phi_{a}^{\ast}(\mathbf{q})} = G^{-1}_{ab}(\mathbf{q}) \, \phi_b ({\mathbf q}) = 0;
\end{equation}
in particular, setting $ \delta F/\delta c (\mathbf{- q}) = 0 $ gives
\begin{equation}
\label{deltac}
\delta c (\textbf{q}) =  \, i \, \frac{(L\, q_{z} - \alpha D \, q_{x})}{D + K_{c} \, q^2} \, u(\mathbf{q}).
\end{equation}

Using (\ref{deltac}) to eliminate the $ \delta c $- field from the free energy (\ref{FFourier}) leads to the effective free energy as a functional of the $ u $- field alone -
\begin{equation}
\label{Feff}
F_{\mathrm{eff}}[u] = \frac{1}{2} \, \frac{B}{\xi^{5}}  \int  \frac{\mathrm{d}^3 p}{(2\pi)^3} \, \frac{g(\mathbf{p})}{1 + \kappa_{c}^{2} \, p^{2}} \, 
u(\mathbf{p})u(-\mathbf{p}) ,
\end{equation}
where the dimensionless wavevector $ \mathbf{p} \equiv \mathbf{q} \, \xi $, and the anisotropic function 
$ g(\mathbf{p}) $ is described in Eqs. (\ref{gp}) and (\ref{g246}) below.
In order to simplify the discussion of stability conditions we have introduced the rescaled dimensionless parameters 
$ l_{B} = \alpha L/B, d_{B} = \sqrt{\alpha^{2} D/B}, \kappa_{u} = \lambda_{u}/\xi,  
\kappa_{c} = \lambda_{c}/\xi $, and the lengths $ \lambda_{u} = \sqrt{K_{u}/B},  \, \lambda_{c} = \sqrt{K_{c}/D} $. In terms of these parameters 
\begin{equation}
\label{gp}
g(\mathbf{p}) = g_{2}(\mathbf{p}) +  g_{4}(\mathbf{p}) +  g_{6}(\mathbf{p}), 
\end{equation}
where
\begin{align}
\label{g246}
g_{2}(\mathbf{p}) &=   [ 1 - (l_{B}/d_{B})^{2} ] \, p_{z}^{2}, \nonumber \\
g_{4}(\mathbf{p}) &= [ \kappa_{u}^{2} \, p^{2} + \kappa_{c}^{2} \{ (p_{z} - d_{B} p_{x} )^{2}  + 2 (d_{B} - l_{B}) p_{x} p_{z} \} ] p^{2}, \nonumber \\
g_{6}(\mathbf{p}) &= \kappa_{u}^{2} \,\kappa_{c}^{2} \, p^{6}.
\end{align}

Thus any equilibrium configuration has to satisfy the condition $ \delta F_{\mathrm{eff}}/\delta u(- \mathbf{p}) = 0 $, which gives 
\begin{equation}
\label{effEL}
\frac{g(\mathbf{p})}{1 + \kappa_{c}^{2} \, p^{2}} \, u(\mathbf{p}) = 0.
\end{equation}

The denominator of the Euler-Lagrange equation (\ref{effEL}) is positive. Therefore it is sufficient to consider the algebraic equation 
$ g(\mathbf{p}) \, u(\mathbf{p}) = 0 $ in analyzing the stability of the Sm$C$ phase. Notice that for the terms constituting $ g(\mathbf{p}) $ the inequalities
\\ (\textit{i}) $ g_{2}(\mathbf{p}) > 0 $ if $ l_{B}^2 < d_{B}^2 $, that is, $ L^{2} < B D $, 
\\ (\textit{ii}) $ g_{4}(\mathbf{p}) > 0 $ if  
$ l_{B}^{2}  <  ( d_{B}^{2} +  (\kappa_{u}/\kappa_{c})^{2} )( 1+ (\kappa_{u}/\kappa_{c})^{2} ) $, and
\\ (\textit{iii}) $ g_{6}(\mathbf{p}) > 0 $ hold for all $ \mathbf{p} $. \\
The above inequalities ensure that that the Euler-Lagrange equation (\ref{effEL}) is satisfied only for $\mathbf{p} = 0$, which corresponds to the Sm$C$ ground state.
Condition (\textit{ii}) is always satisfied if the inequality for the elastic coefficients in (\textit{i}) holds. 
If condition (\textit{i}) does not hold, \textit{i.e.}, if $ L^{2} > B \, D $, and if  $ g_{4}(\mathbf{p}) > 0 $, the equation for stability (\ref{effEL}) has solutions $u(\mathbf{p})$ with nonzero $ \mathbf{p} $.  Thus the range of parameters over which the modulated phase occurs is $ d_{B}^{2} < l_{B}^2 < ( d_{B}^{2} +  (\kappa_{u}/\kappa_{c})^{2} )( 1+ (\kappa_{u}/\kappa_{c})^{2} ) $. 
We note that in the elastic free energy density (\ref{f}) we have not taken into consideration certain symmetry- allowed terms fourth order in $ \mathbf{p} $ and second order in fields (\textit{e.g.}, $ (\nabla^{2} \delta c)^{2} $). Inclusion of such terms broadens the stability range of the modulated phase.

To analyze the modulated phase we use the single- wavevector ansatz
\begin{equation}
\label{uAnsatz}
u(\mathbf{x}) = a \, \cos(q_{x} x + q_{y} y + q_{z} z),
\end{equation}
where $ a $ is the modulation amplitude. The $ \delta c $- field corresponding to this ansatz is given via (\ref{deltac}). 
The average free energy of the modulated phase obtained by using the ansatz (\ref{uAnsatz}) and the corresponding $\delta c$- field, in the free energy (given via (\ref{f}), with the $ K_\phi $- term neglected), and integrating over one spatial period is
\begin{equation}
\label{Fav}
\left\langle f_{\mathrm{eff}} \right\rangle = \frac{A^2 B}{4}  \ \frac{g(\mathbf{p})}{1 + \kappa_{c}^{2} \, p^2},
\end{equation}
with the rescaled modulation amplitude $ A = a/\xi $. Introducing any additional and independent periodic variation in the decoupled field $ \delta \phi $ in the ansatz for the modulated phase increases the average free energy over one period, and is therefore ruled out.

Minimization of the averaged effective free energy (\ref{Fav}) (neglecting the sixth order term in $g(\mathbf{p})$) with respect to  $ \mathbf{p} $ yields the square of the wavenumber
\begin{equation}
\label{psq}
p^{2} \simeq \frac{ (l_{B}/d_{B})^{2} - 1 }{ \kappa_{u}^{2} + (l_{B}/d_{B})^{2} \, \kappa_{c}^{2} },
\end{equation}
and the direction of the wavevector via
\begin{equation}
\label{theta}
\tan 2 \theta \simeq \frac{ 2 \, l_{B} \, \kappa_{c}^{2} }{ \kappa_{u}^{2} + (l_{B}/d_{B})^{2} \, \kappa_{c}^{2} },
\end{equation} 
where $ \tan \theta = p_{x}/p_{z} $. 
The modulation wavevector lies in the $ xz $- plane. This is expected, since the $ L $- term couples distortions in the $ xz $- plane alone. Taking $ K_{u} $ and $ K_{c} $ to be of the order of the Frank elastic constants ($ \sim 10^{-7} $ $ \mathrm{dyne} $), the layer compression modulus $ B \sim 10^7 $ $ \mathrm{dyne \, cm^{-2}} $ \cite{deGP}, and using the fact that the correlation length $ \xi $ is of the order of the smectic layer spacing $ d \sim 10^{-7} $ $ \mathrm{cm} $ \cite{CL}, we get $ \kappa_{c} d_{B} = \alpha \, d^{-1} \sqrt{K_{c}/B} \sim 1 $ and $\kappa_{u} \sim 1$.  Numerical minimization the full, averaged effective free energy (\ref{Fav}) (retaining the sixth order term in $ g(\mathbf{p}) $) with these parameter values gives us the dependence of the components of the dimensionless wavevector on $\mu \equiv (l_{B}^{2}/d_{B}^{2}-1)$ (see \fref{FIG2}). 

Note that the amplitude of modulation $ A $ is governed by higher order (nonlinear) terms in the fields $ \delta c $- and $ u $- in the free energy, and cannot be obtained within the linear theory considered here. However, as an illustrative example, we estimate the amplitude of modulation by including a typical quartic term such as $ C (\partial_{z} u)^4 $ in the elastic free energy density (\ref{f}). Retaining the ansatz  (\ref{uAnsatz}), and the approximate wavevector  given via (\ref{psq}) and (\ref{theta}), minimization of the effective free energy averaged over a period gives $ A \simeq (1/(\eta p_{z}^{2})) \sqrt{ - g(\mathbf{p})/(1 + \kappa_{c}^{2} p^{2})} $, where    
$ \eta  = \sqrt{C/B} $, and we have ignored numerical factors of order unity. 

The condition $ L^{2} > B D $ suggests that the modulated phase can be expected only in materials which are very soft (in that $ D $ is small). The modulated instability is primarily driven by a competition between the elastic constants $ L $, $ D $ and $ B $. However, for small values of $ D $ the penetration depth $ \lambda_{c} $ is likely to be large. It is therefore quite natural that the modulated phase is favored in type-II materials.

Before discussing possible candidate materials for the proposed phase, we point out that: (\textit{i}) In a previous study Johnson and Saupe \cite{JS} found that a material undergoing Sm$A$-Sm$C$ transition exhibited a rectangular grid pattern upon step-by-step cooling across the transition temperature. The Johnson-Saupe instability has two orthogonal wavevectors in the plane of the layers and occurs in cells treated such that the smectic layering is parallel to the cell walls. It is a \textit{metastable} undulation instability which falls in the class of other well known field-induced instabilities such as the Helfrich instability (see, \textit{e.g.}, \cite{deGP}). These are analyzed using nonlinear elasticity-  smectic layers undulate to fill up space as the molecules tilt and the layer spacing reduces. (\textit{ii}) In a sample geometry similar to the one we describe below, modulated structures have been observed in chiral Sm$C^* $ materials. The origin of these have been traced to chiral (and hence polarization) terms in the free energy \cite{Tong, Clark}. (\textit{iii}) Modulated equilibrium structures (ripple phases) are observed in lamellar lyotropic systems on lowering the temperature across the chain- melting transition \cite{KRPR}. Theoretical models for ripple phases are, in essence, based upon Ginzburg-Landau theories exhibiting a Lifshitz point \cite{CLM}. In these models the instability is driven by an elastic coupling between membrane curvature and molecular tilt. The modulated phase proposed in this letter is based upon a novel driving mechanism and gives rise to a \textit{thermodynamically stable structure} with an oblique wavevector in the $ \mathbf{N}_{0} $-$ \mathbf{c}_{0} $- plane.

In what follows, we discuss a previous experiment in which stripe patterns consistent with the proposed structure were observed. We first examine properties of the material used in this experiment. Some dopants are known to enhance the type-II character of mixtures of mesogens. For example, 2-cyano-4-heptylphenyl-4'-pentyl-4-biphenyl carboxylate (7CN5) exhibits the nematic phase with Sm$C$-like (also called skew cybotactic) short-range order over a wide range of temperatures. Adding 7CN5 to a chiral compound exhibiting the Sm$C$$^{*}$ phase induces the TGB$_{A}$ phase, and at a higher concentration, a second, three-dimensionally modulated TGB phase \cite{PPM}. Electroclinic measurements clearly show a rapid decrease in the elastic constant $ D $ with concentration of 7CN5 \cite{DPM}. Indeed, freeze- fracture electron microscopic studies on the three-dimensionally modulated phase demonstrate that the mixture has an extreme type-II character, with Ginzburg parameter $ \sim $ 100, two orders of magnitude larger than that needed for the type-II label \cite{F-etal}. 

Interestingly, experimental studies have also been made on mixtures of an achiral compound exhibiting the Sm$C$ phase with 7CN5 \cite{PHM}. When the mixture is taken in a cell with walls treated for planar alignment of the Frank director $ \mathbf{n} $, the transmitted intensity is crossed out in the nematic phase between appropriately placed crossed polarizers. As the sample is slowly cooled across the two-phase region to the Sm$C$ phase, it develops a stripe pattern oriented along $ \mathbf{n} $. The cell has to be reoriented by $ \pm 1.5^{\circ} $ to get a dark field of view in adjacent stripes, which have a width of about $ 40 \mu $m \cite{PHM,Pthesis}. This observation can be understood if the director pattern of the mixture, which is expected to have a very low value of $ D $, is as shown schematically in \fref{FIG3b}. The wavelength of the observed modulation is $ \sim 80 \mu $m, and the amplitude of tilt- angle modulation is $ \sim 1.5^{\circ} $. This would imply that the deviations from a planar layer structure are quite small.

\end{document}